\documentclass[a4paper,11pt]{article}
\pdfoutput=1
\usepackage{jcappub,xcolor}
\usepackage[T1]{fontenc}
\usepackage[shortlabels]{enumitem}

\begin{document}
\title{Dark energy in light of the early JWST observations: case for a negative cosmological constant?}

\author[a]{Shahnawaz A. Adil,}
\author[b]{Upala Mukhopadhyay,}
\author[b]{Anjan A. Sen,}
\author[c,d]{and Sunny Vagnozzi}
\affiliation[a]{Department of Physics, Jamia Millia Islamia, New Delhi-110025, India}
\affiliation[b]{Centre for Theoretical Physics, Jamia Millia Islamia, New Delhi-110025, India}
\affiliation[c]{Department of Physics, University of Trento, Via Sommarive 14, 38122 Povo (TN), Italy}
\affiliation[d]{Trento Institute for Fundamental Physics and Applications (TIFPA)-INFN, Via Sommarive 14, 38122 Povo (TN), Italy}

\emailAdd{shahnawaz188483@st.jmi.ac.in}
\emailAdd{rs.umukhopadhyay@jmi.ac.in}
\emailAdd{aasen@jmi.ac.in}
\emailAdd{sunny.vagnozzi@unitn.it}

\abstract{Early data from the James Webb Space Telescope (JWST) has uncovered the existence of a surprisingly abundant population of very massive galaxies at extremely high redshift, which are hard to accommodate within the standard $\Lambda$CDM cosmology. We explore whether the JWST observations may be pointing towards more complex dynamics in the dark energy (DE) sector. Motivated by the ubiquity of anti-de Sitter vacua in string theory, we consider a string-inspired scenario where the DE sector consists of a negative cosmological constant (nCC) and a evolving component with positive energy density on top, whose equation of state is allowed to cross the phantom divide. We show that such a scenario can drastically alter the growth of structure compared to $\Lambda$CDM, and accommodate the otherwise puzzling JWST observations if the dynamical component evolves from the quintessence-like regime in the past to the phantom regime today: in particular, we demonstrate that the presence of a nCC (which requires a higher density for the evolving component) plays a crucial role in enhancing the predicted cumulative comoving stellar mass density. Our work reinforces the enormous potential held by observations of the abundance of high-$z$ galaxies in probing cosmological models and new fundamental physics, including string-inspired ingredients.}
\maketitle

\section{Introduction}
\label{sec:introduction}

It is now 25 years since the discovery of cosmic acceleration~\cite{SupernovaSearchTeam:1998fmf,SupernovaCosmologyProject:1998vns}, which led to what is perhaps one of the biggest paradigm shifts in cosmology. While the microphysics underlying cosmic acceleration remains as of yet unknown, the phenomenology thereof is better understood, and usually ascribed to a dark energy (DE) component with negative pressure but positive energy density~\cite{Huterer:2017buf}. In fact, observations are in good agreement with the concordance $\Lambda$CDM model where DE takes the form of a cosmological constant (CC) $\Lambda$, whose microphysical origin is to be sought in the zero-point vacuum energy density of quantum fields~\cite{Carroll:2000fy}. Nevertheless, there are good reasons to believe that a (positive) CC might not be the end of the story: from the theoretical perspective, the expected value of $\Lambda$ is at odds -- to use an euphemism -- with the tiny value required to fit observations~\cite{Weinberg:1988cp}, whereas more recently from the data side, a number of tensions plaguing the $\Lambda$CDM model (e.g.\ tensions affecting the Hubble constant $H_0$~\cite{Verde:2019ivm,Riess:2019qba,DiValentino:2020zio,DiValentino:2021izs,Perivolaropoulos:2021jda,Schoneberg:2021qvd,Shah:2021onj,Abdalla:2022yfr,DiValentino:2022fjm,Hu:2023jqc} and the amplitude of matter fluctuations $S_8$~\cite{DiValentino:2018gcu,DiValentino:2020vvd,Nunes:2021ipq}) might call for new physics in the DE sector. These and other considerations have motivated a wide range of beyond-$\Lambda$ DE models, several of which based on the dynamics of scalar fields (with no claims as to completeness, see e.g.\ Refs.~\cite{Wetterich:1987fm,Ratra:1987rm,Wetterich:1994bg,Caldwell:1997ii,Amendola:1999er,Kamenshchik:2001cp,Capozziello:2002rd,Bento:2002ps,Mangano:2002gg,Farrar:2003uw,Khoury:2003aq,Li:2004rb,Amendola:2006we,Hu:2007nk,Cognola:2007zu,Rinaldi:2014yta,Luongo:2014nld,Myrzakulov:2015qaa,Rinaldi:2015iza,Wang:2016lxa,Josset:2016vrq,Burrage:2016bwy,Sebastiani:2016ras,Nojiri:2017ncd,Burrage:2017qrf,Capozziello:2017buj,Benisty:2018qed,Casalino:2018tcd,Yang:2018euj,Saridakis:2018unr,Langlois:2018dxi,Benisty:2018oyy,Boshkayev:2019qcx,Heckman:2019dsj,DAgostino:2019wko,Mukhopadhyay:2019wrw,Mukhopadhyay:2019cai,Mukhopadhyay:2019jla,Vagnozzi:2019kvw,Akarsu:2019hmw,Saridakis:2020zol,Ruchika:2020avj,Odintsov:2020zct,Odintsov:2020vjb,Oikonomou:2020qah,Oikonomou:2020oex,Vagnozzi:2021quy,Solanki:2021qni,Arora:2021tuh,Narawade:2022jeg,DAgostino:2022fcx,Oikonomou:2022wuk,Belfiglio:2022egm,Kadam:2022yrj,Ong:2022wrs,Belfiglio:2023rxb,Akarsu:2023mfb}).

A feature common to most quintessence scalar field DE models is the positivity of the field potential's ground state, corresponding to a stable or meta-stable de Sitter (dS) vacuum. Interestingly, constructing consistent dS vacua in string theory has proven to be a daunting task, to the point that it has been speculated that string theory may harbor no dS vacua at all~\cite{Danielsson:2018ztv}, or at least no stable ones (as advocated by the swampland program, with important cosmological implications~\cite{Vafa:2005ui,Obied:2018sgi,Agrawal:2018own,Achucarro:2018vey,Garg:2018reu,Kehagias:2018uem,Kinney:2018nny,Ooguri:2018wrx,Palti:2019pca,Grana:2021zvf}). On the other hand anti-de Sitter (AdS) vacua, corresponding to a negative CC (nCC), are ubiquitous within string theory. There are various reasons why this is the case, one of them being the celebrated AdS/CFT correspondence~\cite{Maldacena:1997re}, and more generally the enormous difficulties in formulating a consistent Quantum Field Theory in dS space~\cite{Goheer:2002vf}, from the choice of vacuum state to the (non)-existence of a well-defined S-matrix. In more detail, string theory vacua should be solutions to supergravity theories, and these are typically unstable against quantum/stringy corrections unless they are supersymmetric vacua. There are no known supersymmetric dS vacua in string theory, as the Hamiltonian is a positive-definite operator in any supersymmetric algebra, and dS space does not possess a globally defined time-like Killing vector (along which the Hamiltonian would generate translations)~\cite{Witten:2001kn}. This is the reason why it is easy to construct Freund-Rubin $AdS_p \times S^q$ compactifications~\cite{Freund:1980xh} in $p+q$-dimensional (super)gravity theories with a $q$-form field strength~\cite{DeWolfe:2001nz}. These can be argued to preserve supersymmetry, and thus to be absolutely stable, explaining the ubiquity of AdS vacua in string theory. Constructing parametrically large, long-lived dS vacua has proven extremely difficult in string theory and, if the swampland program is correct, no stable dS vacua should exist.~\footnote{The KKLT~\cite{Kachru:2003aw} and Large Volume Compactification~\cite{Balasubramanian:2005zx} constructions have been argued to provide a mechanism for uplifting AdS vacua into (meta)stable dS vacua, although there is significant debate as to whether the resulting vacua are stable enough.}

The above considerations, along with the fact that negative scalar potentials (thus including AdS vacua) are among the best understood quantum gravity backgrounds by virtue of holography~\cite{Maldacena:1997re}, provide strong motivations for testing the possible existence of a nCC in the DE sector from the cosmological perspective. Of course, a nCC on its own cannot explain cosmic acceleration, so we require an additional component on top, with positive energy density. One can think of such a scenario as describing a scalar field (quintessence) rolling along a potential whose minimum is an AdS vacuum. Here we shall be even more general and allow for an evolving/dynamical DE component (on top of the nCC) which can cross the phantom divide $w_x=-1$, where $w_x$ is the DE equation of state (EoS).~\footnote{Very broadly speaking, one could also argue that such a scenario carries some string motivation. In fact, string compactifications typically predict the existence of a plethora of ultralight (pseudo)scalar particles, which arise from the Kaluza-Klein reduction of higher-dimensional form fields on the topological cycles of the compactification space, in number fixed by the topology of the compactification manifold but typically of order hundreds or more (as in the so-called ``string axiverse'' scenario)~\cite{Svrcek:2006yi,Arvanitaki:2009fg,Cicoli:2012sz,Visinelli:2018utg,Cicoli:2023opf}. While a single standard, minimally coupled scalar field would have an EoS in the quintessence-like regime $w_x>-1$, multiple interacting scalar fields as well as modified gravity scenarios can lead to an \textit{effective} phantom EoS $w_x<-1$~\cite{Carroll:2003st,Vikman:2004dc,Carroll:2004hc,Deffayet:2010qz,Sawicki:2012pz}. Broadly speaking, this motivates studying a scenario featuring an evolving DE component, which can potentially cross the phantom divide, on top of a nCC.} Such a scenario has been studied earlier in light of various cosmological datasets~\cite{Cardenas:2002np,Dutta:2018vmq,Visinelli:2019qqu,Calderon:2020hoc,Sen:2021wld},~\footnote{See also Ref.~\cite{Malekjani:2023dky} for related results, as well as Refs.~\cite{Ye:2020btb,Ye:2020oix,Ye:2021nej,Jiang:2021bab,Ye:2021iwa} for recent works on cosmological implications of an AdS phase in the early Universe, specifically in the context of early dark energy~\cite{Poulin:2023lkg}.} and our goal is to open a new window onto this class of models using observations of high-$z$ galaxies.

Although most DE tests are carried out using low-$z$ ($z \lesssim 2$), standard cosmological probes, it is worth studying whether other probes in different redshift ranges may help further shed light on this elusive component. A potentially interesting probe in this sense is the abundance of high-$z$ massive galaxies: these populate high-mass dark matter (DM) haloes, whose mass function and redshift evolution thereof depend strongly on the underlying cosmology, in particular on the background expansion and growth rate, both of which can potentially be drastically altered \textit{even at high redshift} within alternative DE models. The \textit{predicted} abundance of massive DM haloes at a given redshift can be compared to the \textit{observed} abundance of galaxies of given stellar mass in a conservative way bypassing complications introduced by complex galaxy formation physics. In fact, the stellar mass content $M_{\star}$ of a galaxy is bounded above by the galaxy's total baryonic matter, whose maximum is determined if the DM halo mass $M$ and cosmic baryon fraction $f_b \equiv \Omega_b/\Omega_m$ are known: $M_{\star} \leq f_bM$.~\footnote{In general this relation is more properly written as $M_{\star}=\epsilon f_bM$, where $\epsilon$ is the efficiency of converting gas into stars, whose value is expected to be $\epsilon \lesssim 0.2$, with a moderate redshift dependence~\cite{Leroy:2008kh,Combes:2010vc,Tacchella:2018qny}.} As galaxies cannot outnumber their DM haloes, the observed abundance of high-$z$ galaxies of known stellar mass can be used to exclude cosmological models which do not allow for a sufficiently rapid growth of DM haloes, potentially including various DE models, as pointed out in Ref.~\cite{Menci:2020ybl}.

Such a comparison has become all the more urgent in light of early observations from the long-awaited, next-generation space telescope: the James Webb Space Telescope (JWST)~\cite{Gardner:2006ky}, which has given us a first glimpse of high-$z$ galaxy formation. Initial JWST imaging data from NIRCam observations of the Cosmic Evolution Early Release Science (CEERS) program have in fact uncovered the existence of a surprisingly abundant (cumulative stellar mass density $\rho_{\star} \gtrsim 10^6M_{\odot}/{\rm Mpc}^3$) population of massive galaxies ($M_{\star} \gtrsim 10^{10.5}M_{\odot}$) at very high redshift ($7 \lesssim z \lesssim 10$)~\cite{Naidu:2022ghw,Castellano:2022ghw,Adams:2022ghw,Atek:2022ghw,Labbe:2023ghw}. While these results come with important caveats (e.g.\ the redshifts are photometric and only a handful have been spectroscopically confirmed~\cite{Bouwens:2022gqg,Xiao:2023ghw}), should they hold up to scrutiny, the challenge they would pose to the current concordance $\Lambda$CDM model would be huge~\cite{Boylan-Kolchin:2022kae}, just as the results from JWST's predecessor, the Hubble Space Telescope, challenged the then-current and eventually dethroned Einstein-dS model~\cite{Bolte:1995ghw,Krauss:1995yb,Ostriker:1995su}: in fact, several works have already started exploring the intriguing possibility that the early JWST results may be calling for new fundamental physics, or more generally the potential of early JWST data to constrain fundamental physics, see e.g.\ Refs.~\cite{Biagetti:2022ode,Haslbauer:2022vnq,Hutsi:2022fzw,Gandolfi:2022bcm,Maio:2022lzg,Lovyagin:2022kxl,Wang:2022jvx,Yuan:2023bvh,Dayal:2023nwi,Ilie:2023zfv,Jiao:2023wcn,Parashari:2023cui,Hassan:2023asd,Lei:2023mke,Yoshiura:2023xkd,John:2023eck,Padmanabhan:2023esp,Lin:2023ewc,Su:2023jno,Forconi:2023izg,Gouttenoire:2023nzr,Guo:2023hyp,Huang:2023chx,Bird:2023pkr,Wang:2023ros,Gupta:2023mgg}.~\footnote{Of course, the JWST results may also imply our need to rethink the physics of galaxy formation, see e.g.\ Refs.~\cite{Ferrara:2022dqw,Qin:2023rtf,Pallottini:2023yqg,Wang:2023xmm,Pacucci:2023oci} for works in this direction.}

In this paper we shall take a more conservative approach, and assess the potential of the JWST observations in discriminating between different DE models, and in particular in testing the possible existence of a nCC in the DE sector. In fact it was recently argued by Ref.~\cite{Menci:2022wia} that, if taken seriously, the early JWST observations rule out a major portion of the parameter space of dynamical DE models currently consistent with cosmological data. The implications of these conclusions can be very far-reaching. Here we show how these conclusions can be drastically altered if a nCC is introduced in the dark sector. Our results open a new window towards testing an important prediction of string theory using observations of massive, high-$z$ galaxies, with minimal assumptions on galaxy formation.

The rest of this paper is then organized as follows. In Sec.~\ref{sec:models} we introduce the nCC and discuss the DE models we consider. In Sec.~\ref{sec:highredshiftobservables} we discuss the methods used to compute the observables relevant for high-$z$ galaxies. In Sec.~\ref{sec:jwst} we then compare our predicted cumulative stellar mass densities to the abundance of high-$z$ galaxies inferred from JWST observations. Finally, we draw closing remarks in Sec.~\ref{sec:conclusions}.

\section{Dark energy models}
\label{sec:models}

We now present the dark energy models considered, some of which include a nCC, while others do not, in order to provide a comparative assessment of the effect of a nCC on the observables studied. For what concerns the DE sector, we do not commit to any specific microphysical model for DE, but rather adopt a general parametrization for an evolving DE EoS $w_x(z)$. Specifically, we shall consider the widely used Chevallier-Polarski-Linder (CPL) parametrization for $w_x(z)$, given by~\cite{Chevallier:2000qy,Linder:2002et}:
\begin{eqnarray}
w_x(z)=w_0 + w_a\frac{z}{1+z}\,,
\label{eq:cpl}
\end{eqnarray}
where $z$ denotes redshift, and $w_0$ and $w_a$ are two constants, with $w_0$ corresponding to the present-time DE EoS. Clearly, Eq.~(\ref{eq:cpl}) corresponds to a Taylor series of the DE EoS in powers of the scale factor $a=(1+z)^{-1}$ around the present value $a_0=1$, truncated to first order. There are several reasons why the CPL parametrization is widely used. Firstly, its 2-dimensional nature makes it easy to manage from the computational point of view. Next, and perhaps most importantly, it has a direct connection to several physical DE models, notably quintessence DE. In fact, after being tested against physical solutions of the Klein-Gordon equation, the CPL parametrization has been shown to be accurate to sub-percent level in recovering observables (e.g.\ expansion rate and distance measurements) for quintessence DE models for different choices of $w_0$ and $w_a$~\cite{Linder:2002et,Linder:2002wx,Linder:2006sv,Linder:2007wa,Linder:2008pp,Scherrer:2015tra}. The only other parametrization which has been shown to possess these properties is the 4-parameter Copeland-Corasaniti-Linder-Huterer parametrization~\cite{Corasaniti:2002vg,Linder:2005ne}. While a number of other EoS parametrizations for dynamical DE have been proposed in the literature,~\footnote{Most of these have been proposed with phenomenological motivations, see e.g.\ Refs.~\cite{Efstathiou:1999tm,Jassal:2004ej,Gong:2005de,Barboza:2008rh,Ma:2011nc,Pantazis:2016nky,Yang:2017alx,Pan:2017zoh,Yang:2018qmz,Singh:2023ryd} for examples in this sense (see also Ref.~\cite{Perkovic:2020mph} for a study on the theoretical viability conditions for these parametrizations, and Ref.~\cite{Colgain:2021pmf} for discussions on potential shortcomings of the CPL parametrization), although the community has been moving towards non-parametric reconstructions of the DE EoS and related quantities, see e.g.\ Refs.~\cite{Seikel:2012uu,Hee:2016nho,Shafieloo:2018gin,Wang:2018fng,Gerardi:2019obr,DiazRivero:2019ukx,Bonilla:2020wbn,Dhawan:2021mel,Dialektopoulos:2021wde,Bernardo:2022pyz}.} here we shall stick with the CPL parametrization given its appealing properties discussed above. Note that at asymptotically early times, i.e.\ for $z \to \infty$, the CPL EoS tends to $w_0+w_a$. Therefore, even if the DE component is presently in the quintessence-like regime ($w_0>-1$), if $w_a$ is sufficiently negative the DE component will eventually become phantom in the past - similar considerations of course hold for a DE component which is phantom at present but can become quintessence-like in the past. For a dynamical DE model described by the CPL parametrization, the energy density $\rho_x(z)$ evolves as follows:
\begin{eqnarray}
\rho_x(z)=\Omega_x\rho_{\rm crit}^{(0)}(1+z)^{3(1+w_0+w_a)}\exp \left ( -3w_a \dfrac{z}{1+z} \right ) \,,
\label{eq:rhocpl}
\end{eqnarray}
where $\rho_{\rm crit}^{(0)}$ and $\rho_x^{(0)}$ are the current critical energy density and (positive) energy density of the CPL component respectively, and $\Omega_x \equiv \rho_x^{(0)}/\rho_{\rm crit}^{(0)}$ is the CPL DE density parameter.

On top of the DE component described by Eq.~\ref{eq:cpl} we also consider the presence of a nCC in the dark sector, $\Lambda<0$, with associated nCC density parameter $\Omega_{\Lambda}=\Lambda/3<0$. Considering also the presence of radiation and matter with density parameters $\Omega_r$ and $\Omega_m$ respectively, and assuming a spatially flat Universe, the first Friedmann equation governing the evolution of the Hubble rate can be expressed in the following form:
\begin{eqnarray}
H^2(z)=H_0^2 \left [ \Omega_r(1+z)^4+\Omega_m(1+z)^3+\Omega_{\Lambda}+\Omega_x(1+z)^{3(1+w_0+w_a)}\exp \left ( -3w_a \frac{z}{1+z} \right) \right ] \,,\nonumber \\
\label{eq:hubblerate}
\end{eqnarray}
where $H_0$ is the Hubble constant. As we are assuming a spatially flat Universe, the density parameters are not independent, but obey the constraint:
\begin{eqnarray}
\Omega_r+\Omega_m+\Omega_{\Lambda}+\Omega_x = 1\,.
\label{eq:flat}
\end{eqnarray}
Moreover, it is natural to identify the \textit{combined} DE sector as comprising the CPL component with density parameter $\Omega_x$, and the nCC component with density parameter $\Omega_{\Lambda}$, and therefore we make the following identification:
\begin{eqnarray}
\Omega_{\rm DE} = \Omega_x+\Omega_{\Lambda} \simeq 1-\Omega_m\,,
\label{eq:omegade}
\end{eqnarray}
where the last equality is approximately true when neglecting the radiation component, completely subdominant today, whereas we can identify the total DE energy density as being:
\begin{eqnarray}
\rho_{\rm DE}(z) = \rho_x(z)+\rho_{\Lambda}=\Omega_x\rho_{\rm crit}^{(0)}(1+z)^{3(1+w_0+w_a)}\exp \left ( -3w_a \dfrac{z}{1+z} \right )+\Omega_{\Lambda}\rho_{\rm crit}^{(0)}\,,
\label{eq:rhode}
\end{eqnarray}
Note that it is $\Omega_{\rm DE}$, i.e.\ the sum of the dynamical DE and nCC components, which is required to be $\approx 0.7$ in order to satisfy current observational constraints on the amount of DE. Therefore, observations approximately fix the sum $\Omega_x+\Omega_{\Lambda}$, but the two terms can in principle take any value (though $\Omega_{\Lambda}$ itself is subject to order unity constraints from cosmology). Moreover, it is this combined DE sector which has to be able to account for cosmic acceleration. From very general considerations, if $\Lambda$ takes more negative values (i.e.\ $\vert \Omega_{\Lambda} \vert$ is increased), then more negative values of $w_x$, i.e.\ a dynamical DE component moving towards the phantom regime, are required in order for cosmic acceleration to occur today, in agreement with observations. In general, different combinations of $\Omega_x$ and $\Omega_{\Lambda}$ can lead to a rich phenomenology in the combined DE sector, including periods of transient acceleration potentially followed by a period of future deceleration~\cite{Calderon:2020hoc}.

Note, in addition, that we can define an effective EoS associated to the total DE sector, which is essentially given by a weighted average of the nCC (with EoS $-1$) and positive DE (with CPL EoS) components. By using the continuity equation, itself following from local energy conservation as dictated by the twice contracted Bianchi identity, the effective equation of state $w_{\rm eff}(z)$ of the DE sector, with total energy density $\rho_{\rm DE}$ given by Eq.~(\ref{eq:rhode}), can be written as:
\begin{eqnarray}
w_{\rm eff}(z) &=& \frac{1}{3}\frac{d\ln\rho_{\rm DE}(z)}{d\ln(1+z)}-1 = \frac{1+z}{3\rho_{\rm DE}(z)}\frac{d\rho_{\rm DE}(z)}{dz}-1 \nonumber \\
&=&\frac{\Omega_x(1+z)^{2+3w_0+3w_a} \left [ w_0+(w_0+w_a)z \right ] \exp \left ( -3w_a \dfrac{z}{1+z} \right ) -\Omega_{\Lambda}}{\Omega_x(1+z)^{3(1+w_0+w_a)}\exp \left ( -3w_a \frac{z}{1+z} \right ) +\Omega_{\Lambda}}
\label{eq:weff}
\end{eqnarray}
which reduces to $(w_0\Omega_x-\Omega_{\Lambda})/(\Omega_x+\Omega_{\Lambda})$ as $z \to 0$, showing that the present-day effective EoS of the DE sector is $w_{\rm eff}(z=0) \neq w_0$ unless $\Omega_{\Lambda}=0$. It is easy to show that Eq.~(\ref{eq:weff}) can also be obtained by taking the ratio between the total pressure of the DE sector and its total energy density, where the latter is given by Eq.~(\ref{eq:rhode}), and the nCC contributes with a positive pressure term $P_{\Lambda}=-\Omega_{\Lambda}\rho_{\rm crit}^{(0)}>0$.

In the following, our aim is to investigate the effect of a nCC on high-$z$ galaxies, such as those observed by the JWST. In order to do so, we shall consider 9 different models, corresponding to 9 different choices of the CPL parameters $w_0$ and $w_a$, and the nCC parameter $\Omega_{\Lambda}$, appearing in Eq.~(\ref{eq:hubblerate}). The 9 models we consider feature different combinations of phantom and quintessence-like DE components, with and without phantom crossing and/or a nCC component, and are given as follows:
\begin{enumerate}[1)]
\item $\Lambda$CDM model (positive CC): $w_0=-1$, $w_a=0$, $\Omega_\Lambda=0$, $\Omega_x=0.7$;
\item quintessence-like DE ($w_x>-1$) with constant DE EoS and no nCC: $w_0=-0.95$, $w_a=0$, $\Omega_\Lambda=0$, $\Omega_x=0.7$;
\item phantom DE ($w_x<-1$) with constant DE EoS and no nCC: $w_0=-1.05$, $w_a=0$, $\Omega_\Lambda=0$, $\Omega_x=0.7$;
\item dynamical DE which is quintessence-like today ($w_x>-1$), but was phantom in the past, with no nCC: $w_0=-0.95$, $w_a=-1$, $\Omega_\Lambda=0$, $\Omega_x=0.7$;
\item dynamical DE which is phantom today ($w_x<-1$), but was quintessence-like in the past, with no nCC: $w_0=-1.05$, $w_a=1$, $\Omega_\Lambda=0$, $\Omega_x=0.7$;
\item as in case 2), but with an additional nCC: $w_0=-0.95$, $w_a=0$, $\Omega_\Lambda=-1.5$, $\Omega_x=2.2$;
\item as in case 3), but with an additional nCC: $w_0=-1.05$, $w_a=0$, $\Omega_\Lambda=-1.5$, $\Omega_x=2.2$;
\item as in case 4), but with an additional nCC: $w_0=-0.95$, $w_a=-1$, $\Omega_\Lambda=-1.5$, $\Omega_x=2.2$;
\item as in case 5), but with an additional nCC: $w_0=-1.05$, $w_a=1$, $\Omega_\Lambda=-1.5$, $\Omega_x=2.2$.
\end{enumerate}
Note that in all cases, the total present-day density parameter of the DE sector [$\Omega_{\rm DE}$ given in Eq.~(\ref{eq:omegade})] is $0.7$, resulting from a combination of a nCC and a positive evolving CPL DE density. Finally, we note that models such as 4), 5), 8), and 9), featuring crossing between the phantom and quintessence-like regimes (in either direction) have also been referred to as ``quintom'' models in the literature~\cite{Feng:2004ff,Guo:2004fq,Cai:2007gs,Cai:2007qw,Zhang:2009un,Saridakis:2009ej,Cai:2012yf,Bahamonde:2018miw,Leon:2018lnd,Panpanich:2019fxq}, see e.g. Ref.~\cite{Cai:2009zp} for a review.
\begin{figure}[!t]
\centering
\includegraphics[width=0.49\linewidth]{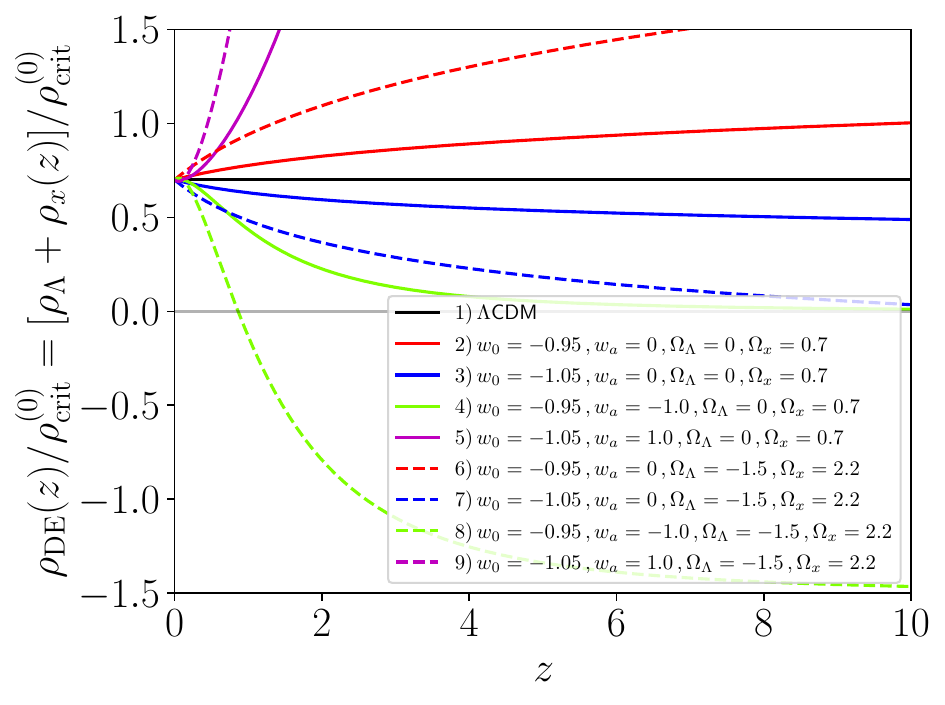}\,
\includegraphics[width=0.49\linewidth]{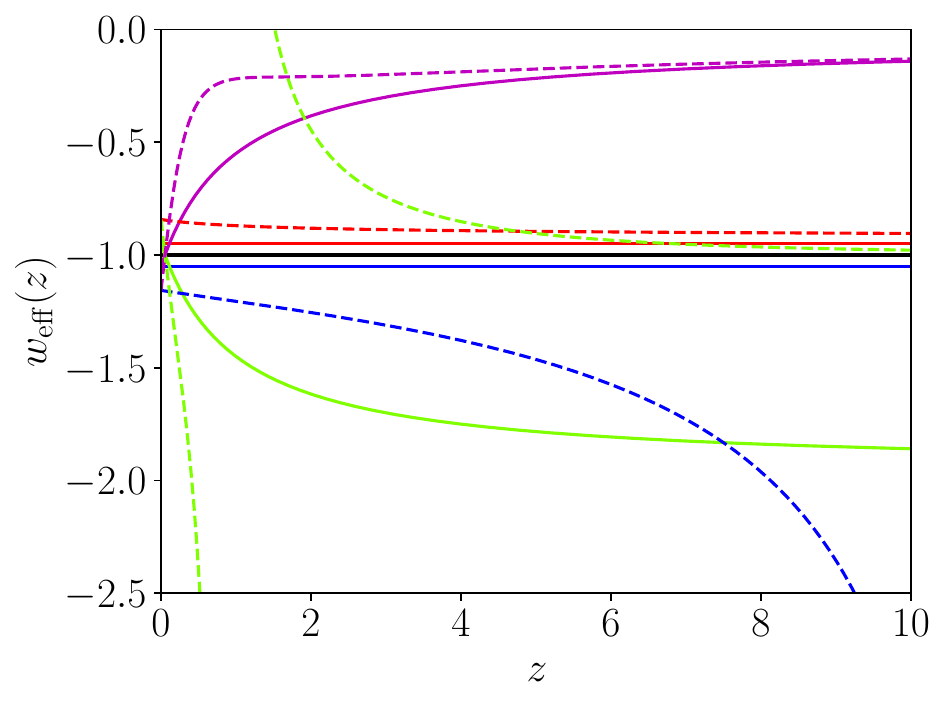}
\caption{\textit{Left panel}: evolution of the total dark density (negative cosmological constant plus evolving dark energy) as a function of redshift, given by Eq.~(\ref{eq:rhode}) and divided by the critical density, for the 9 different dark energy models summarized in Sec.~\ref{sec:models}, as determined by the color coding. \textit{Right panel}: evolution of the effective equation of state of the dark energy sector as a function of redshift, given by Eq.~(\ref{eq:weff}), for the 9 different dark energy models summarized in Sec.~\ref{sec:models}, and with the same color coding as in the left panel. The appearance of poles in the effective equation of state for models 7) and 8) is a direct consequence of the corresponding energy densities changing sign at a certain redshift, as is clear from the left panel (the horizontal grey line corresponds to $\rho_{\rm DE}=0$).}
\label{fig:rhow}
\end{figure}

In Fig.~\ref{fig:rhow}, we show the redshift evolution of the energy density and effective EoS of the total DE component, given by Eq.~(\ref{eq:rhode}) and Eq.~(\ref{eq:weff}) respectively, for the 9 models described above. Note that for models 7) and 8), featuring a nCC and respectively a phantom DE component and a quintom DE component (phantom in the past), the total DE energy density changes sign at sufficiently high redshift. In particular, for the choice of parameters given above, the sign switch occurs at redshift $z_p \sim 0.86$ for model 7), and at $z_p \sim 11.85$ for model 8), although the latter is not visible in the Figure. This sign change is reflected in the presence of a pole in the corresponding effective EoS (hence the subscript ``\textit{p}'' in the pole redshift $z_p$), which diverges to negative infinity as $z \to z_p^{-}$, and to positive infinity as $z \to z_p^{+}$. This divergence does not signal a pathology, as the quantity which diverges is only the \textit{effective} EoS of the total DE sector (nCC plus CPL DE component), which is an useful effective quantity but cannot otherwise be associated to a single fundamental microscopical degree of freedom. More importantly, the appearance of a pole in the effective EoS of a sign-switching perfect fluid component in a spatially uniform Universe has recently been rigorously demonstrated to be a consequence of local energy conservation, regardless of the underlying sources which are responsible for the sign switch~\cite{Ozulker:2022slu}. As such, the appearance of poles in the green dotted and blue dotted curves (the latter not visible given the plotted redshift range) of Fig.~\ref{fig:rhow} should not come as a surprise. Rather they show the importance, especially in the context of non-parametric reconstructions, of focusing on the total DE energy density rather than its effective EoS, as stressed in Ref.~\cite{Ozulker:2022slu}.

\section{High-redshift galaxy observables}
\label{sec:highredshiftobservables}

We now discuss the methods used to compute the observables relevant for high-$z$ galaxies, which include the structure growth rate, as well as comoving (cumulative) number or mass (stellar mass or DM halo mass) densities.

\subsection{Density perturbations and growth factor}
\label{subsec:densityperturbationsgrowth}

To account for the effects on high-$z$ galaxies of the different DE models considered, described in Sec.~\ref{sec:models}, the first step is to compute the evolution of the matter density contrast $\delta(a)$. On sub-horizon scales, i.e.\ those relevant for galaxy formation, and assuming that DE does not cluster, the evolution of $\delta$ is governed by the following growth equation:
\begin{eqnarray}
\delta^{\prime\prime} + \left ( \frac{3}{a}+\frac{E^\prime}{E} \right ) \delta^{\prime} -\frac{3}{2}\frac{\Omega_m}{a^5 E^2}\delta=0\,,
\label{eq:growth}
\end{eqnarray}
where $^\prime$ indicates a derivative with respect to the scale factor $a$, and $E(a) \equiv H(a)/H_0$ denotes the normalized expansion rate. Therefore, besides modifying the background expansion rate given by Eq.~(\ref{eq:hubblerate}), the different DE models we consider also affect the growth of structure by altering $E(a)$ in Eq.~(\ref{eq:growth}).

\begin{figure}[!t]
\centering
\includegraphics[scale=0.8]{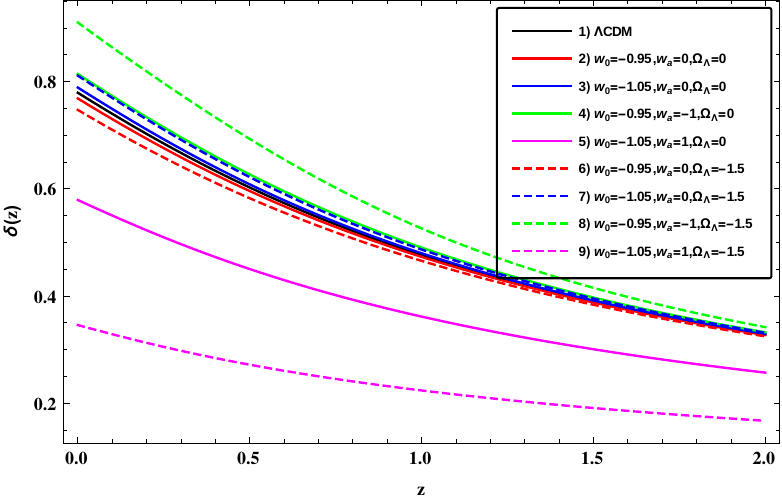}
\caption{Evolution of the linear matter density contrast as a function of redshift, for the 9 different dark energy models summarized in Sec.~\ref{sec:models}, with the same color coding as in Fig.~\ref{fig:rhow}.}
\label{fig:growth}
\end{figure}

We solve Eq.~(\ref{eq:growth}) numerically for the 9 DE models presented in Sec.~\ref{sec:models}, setting initial conditions $\delta(a_i) \sim a$ and $\delta'(a_i) \sim 1$ at some initial scale factor $a_i \ll 1$ deep in the matter-dominated era. The result is shown in Fig.~\ref{fig:growth}, where we plot $\delta(z)$. From this Figure, taking as reference the black curve corresponding to $\Lambda$CDM, we notice a few interesting points.
\begin{itemize}
\item Whether a DE model leads to larger or smaller density contrasts relative to $\Lambda$CDM appears to be driven by whether the models are quintessence-like or phantom \textit{in the past} rather than today. In fact, neglecting for the moment the nCC, we see that the two models which lead to larger density contrasts relative to $\Lambda$CDM are model 3) with constant phantom EoS (blue curve), and model 4) which was phantom in the past, but crossed the phantom divide and is quintessence-like today (green curve). In contrast, the two models which lead to smaller density contrasts relative to $\Lambda$CDM are model 2) with constant quintessence-like EoS (red curve), and model 5) which was quintessence-like in the past, but crossed the phantom divide and is phantom today (purple curve).
\item Adding a nCC exacerbates all the previous effects: to see this, compare solid versus dashed curves for a fixed color. In other words, models which previously led to larger density contrasts relative to $\Lambda$CDM now see their enhancement effect amplified [3) $\to$ 7) and 4) $\to$ 8)], and conversely models which previously led to smaller density contrasts relative to $\Lambda$CDM see their suppression effect amplified [2) $\to$ 6) and 5) $\to$ 9)].
\end{itemize}
Overall, the inclusion of a nCC therefore amplifies in either direction the effects of dynamical DE on structure growth.

The previous findings can be understood as follows. When looking back in time, or equivalently going to higher redshift, and keeping all other parameters fixed, phantom models lead to a lower expansion rate relative to $\Lambda$CDM (or equivalently a stronger acceleration as one moves towards lower redshifts). This suppression in $E(z)$ results in an enhancement of the driving term $\propto 1/E^2$ appearing in Eq.~(\ref{eq:growth}), explaining why the effect of phantom DE is to enhance the growth of structure, in agreement with previous findings. These effects are exacerbated by the inclusion of a nCC simply because, by virtue of Eq.~(\ref{eq:omegade}), including a nCC requires increasing $\Omega_x$ in order for $\Omega_{\rm DE} \sim 0.7$ to hold: to put it differently, a nCC opposes cosmic acceleration, and therefore one requires a larger positive evolving DE density to counteract this effect and remain in agreement with observations. This therefore enhances the DE effects because the energy density of the evolving component, $\Omega_x$, is increased. In some sense, the effect of the nCC on the growth of structure is therefore indirect, resulting from a parameter shift enforced by Eq.~(\ref{eq:omegade}) and, ultimately, by Eq.~(\ref{eq:flat}).~\footnote{It is worth mentioning that we have also calculated $\delta_c$, the critical density contrast for a linear overdensity at the redshift of collapse, using Eq.~(\ref{eq:growth}) and following Refs.~\cite{Pace:2010sn,Campanelli:2011qd,Mota:2008ne}. Specifically, we have first solved the full non-linear equation for the evolution of the density contrast and found the initial condition for which spherical collapse or divergence of $\delta$ occurs at $z\sim 7$. We have then solved the linear equation for the evolution of the density contrast, and used this to determine $\delta_c$. We have found that for all DE models described in Sec.~\ref{sec:models}, $\delta_c \simeq 1.686$ holds at high redshift.}
\begin{figure}[!t]
\centering
\includegraphics[width=0.49\linewidth]{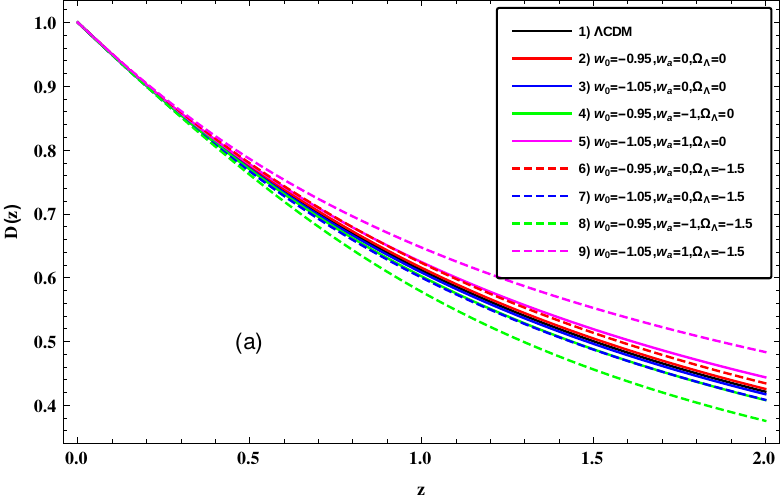}\,
\includegraphics[width=0.49\linewidth]{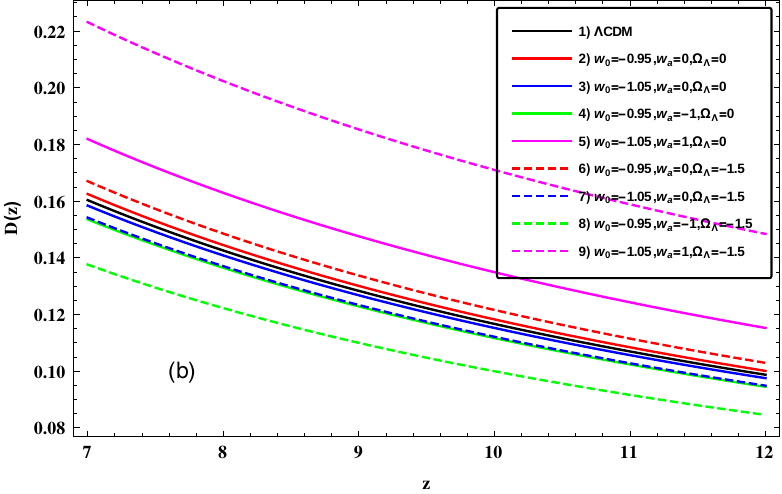}
\caption{Evolution of the growth factor as a function of redshift, for the 9 different dark energy models summarized in Sec.~\ref{sec:models}, with the same color coding as in Fig.~\ref{fig:rhow}. The evolution is shown for both low redshifts [$0 \leq z \leq 2$, left panel (a)] and high redshifts [$7 \leq z \leq 12$, right panel (b)].}
\label{fig:growthfactor}
\end{figure}

One further important quantity is the growth factor, which in linear perturbation theory is defined as:
\begin{eqnarray}
D(z) \equiv \frac{\delta(z)}{\delta(0)}\,,
\label{eq:dz}
\end{eqnarray}
and can therefore be calculated from Eq.~(\ref{eq:growth}). Once we have solved Eq.~(\ref{eq:growth}), we therefore also calculate $D(z)$, with the results shown in Fig.~\ref{fig:growthfactor} for the same models considered earlier, and in two different redshift ranges: low [$z \lesssim 2$, left panel (a)] and high [$7 \lesssim z \lesssim 12$, right panel (b)] redshift. Here we find a completely opposite trend relative to that seen in Fig.~\ref{fig:growth}: in other words, models 3) and 4) which were phantom in the past and led to larger density contrasts relative to $\Lambda$CDM, in this case lead to a lower growth factor -- conversely, models 2) and 5) which were quintessence-like in the past and led to smaller density contrasts relative to $\Lambda$CDM, in this case lead to a larger growth factor. Again, however, we see that adding a nCC exacerbates all these effects (once more, compare solid versus dashed curves for a fixed color). While perhaps somewhat counterintuitive, this effect can be understood from the fact that $\delta(0)$ appears in the denominator of Eq.~(\ref{eq:dz}). The reason why a nCC exacerbates these effects remains the same as the one discussed earlier, and is therefore somewhat indirect, as a result of the required increase in $\Omega_x$ once a negative $\Omega_{\Lambda}$ is introduced.

\begin{figure}[!t]
\centering
\includegraphics[scale=0.7]{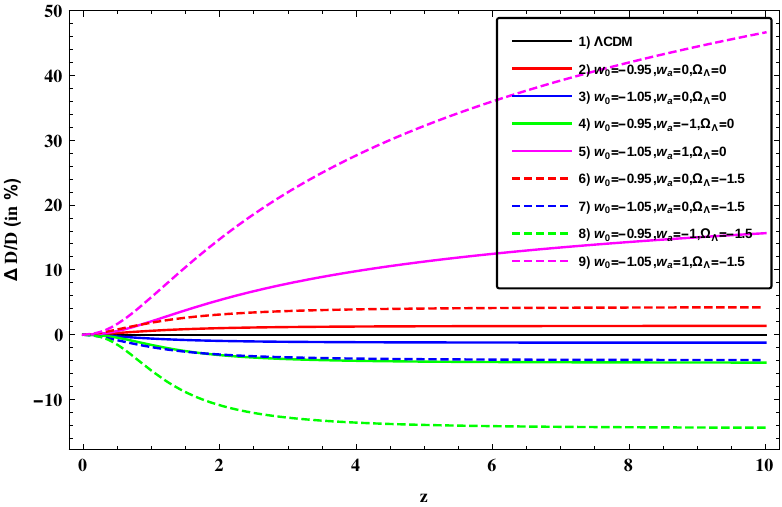}
\caption{Percent relative deviation between the growth factors of the different dark energy models [$100 \times (D(z)_{\rm DE}-D(z)_{\Lambda{\rm CDM}})/D(z)_{\Lambda{\rm CDM}}$], for the 9 different dark energy models summarized in Sec.~\ref{sec:models}, with the same color coding as in Fig.~\ref{fig:rhow}.}
\label{fig:percent}
\end{figure}

By construction, all the growth factors are equal to each other at $z=0$: $D(0)=1$. In contrast, we see that the growth factors for the different DE models can deviate significantly from the $\Lambda$CDM evolution at high redshift, as the lower panel of Fig.~\ref{fig:growthfactor} clearly shows for the redshift range $7 \lesssim z \lesssim 12$. Overall, the model which most strongly enhances the growth of structure at high redshift is therefore model 9), which features a nCC, was quintessence-like in the past, but crossed the phantom divide at some point and is therefore phantom today (dashed purple curves). Qualitatively, we can therefore already expect this model to lead to the strongest enhancement in the abundance of high-$z$ galaxies, of interest for the JWST observations given the considerations made in Sec.~\ref{sec:introduction}: this expectation will be confirmed quantitatively in Sec.~\ref{sec:jwst}. To showcase these effects more clearly, in Fig.~\ref{fig:percent} we plot the (percent) relative deviation between the growth factors of the different dark energy models: $\Delta D/D \equiv (D(z)_{\rm DE}-D(z)_{\Lambda{\rm CDM}})/D(z)_{\Lambda{\rm CDM}}$. Here we clearly see that the largest deviation is for the phantom-to-quintessence-like-crossing (as time goes on) model with a nCC, whose relative deviation with respect to $\Lambda$CDM increases with increasing redshift, before saturating at $\lesssim 50\%$ at redshift $z \sim 10$. We see that other models lead to a smaller relative growth (or suppression), and that in all cases the relative deviation with respect to $\Lambda$CDM saturates at sufficiently high redshift.

\subsection{Comoving number and mass densities}
\label{subsec:densities}

We now proceed to compute comoving (cumulative) number or mass (stellar mass or DM halo mass) densities for the different models considered. To compute these quantities we adopt the Sheth-Tormen (ST) prescription~\cite{Sheth:1999mn}, itself extending the Press-Schechter (PS) formalism~\cite{Press:1973iz}, which we briefly review.

The DM halo mass function $dn(M,z)/dM$ quantifies the number of DM haloes of mass $M$ per unit mass per unit co-moving volume at
redshift $z$ in the mass range $[M;M+dM]$, and is given by:
\begin{eqnarray}
\frac{dn}{dM}=-\frac{\rho_m^{(0)}}{M}\frac{d{\rm ln}\sigma}{dM}f(\sigma)\,,
\label{eq:dndm}
\end{eqnarray}
where $\rho_m^{(0)}$ is the present-day matter energy density, $\sigma$ is the variance of matter density fluctuations in a sphere of comoving radius $R$, and $f(\sigma)$ will be defined later. The mass $M$ within the sphere of comoving radius $R$ is given by $M=4\pi R^3 \rho_m^{(0)}/3$, whereas $\sigma$, which is related to the power spectrum of matter density perturbations $P(k,z)$, is defined by:
\begin{eqnarray}
\sigma^2(R,z)=\frac{1}{2\pi^2}\int_0^\infty dk\,k^2P(k,z)W^2(kR)\,.
\label{eq:sigma}
\end{eqnarray}
In the above, we have implicitly moved to Fourier space, where $k$ denotes the perturbation wavenumber, and $W(kR)$ is a window function used to smooth the density field on a certain scale. Here we consider a top-hat filter in real space, whose Fourier transform is given by $W(x) = 3(\sin x - x\cos x)/x^3$. Finally, in Eq.~(\ref{eq:sigma}), $P(k,z)$ denotes the power spectrum of matter density fluctuations, which is given by:
\begin{eqnarray}
P(k,z)=P_0(k) T^2(k) D^2(z)\,,
\label{eq:pk}
\end{eqnarray}
with $P_0(k)$ a normalization related to the present-day value of $\sigma_8$, whereas $T(k)$ is the transfer function and $D(z)$ is the linear growth factor which we discussed earlier. We use the ST prescription~\cite{Sheth:1999mn}, which extended the PS formalism to account for ellipsoidal (rather than spherical) collapse, and for which $f(\sigma)$ is given by:
\begin{eqnarray}
f(\sigma)=A \sqrt{\frac{2a}{\pi}} \left [ 1+ \left ( \frac{\sigma^2}{a\delta_c^2} \right ) ^p \right ] \frac{\delta_c}{\sigma}\exp \left ( -\frac{\delta_c^2 a}{2\sigma^2} \right ) \,,
\label{eq:st}
\end{eqnarray}
where the three parameters are given by $A=0.322$, $a=0.707$, and $p=0.3$ respectively, and the PS mass function is recovered in the limit $a \to 1$ and $p \to 0$.~\footnote{Although in principle high-resolution N-body simulations may provide better fitting functions compared to ST mass function~\cite{Basilakos:2009mz,Bhattacharya:2010wy}, here we have opted for using the ST mass function as we are only interested in providing a first assessment of the potential viability of DE models featuring a nCC in light of JWST data, leaving a more refined analysis to future work, should these models be found to be interesting in this context.}

Finally, the quantity of interest to high-redshift galaxy observations is the cumulative comoving DM halo number density $n(>M_{\rm halo},z)$, which quantifies the number density of haloes above a given mass threshold $M_{\rm halo}$, and is given by~\cite{Boylan-Kolchin:2022kae}:
\begin{eqnarray}
n(>M_{\rm halo},z)=\int_{M_{\rm halo}}^\infty dM\,\frac{dn(M,z)}{dM}\,.
\label{eq:nhalo}
\end{eqnarray}
A closely related quantity is the cumulative halo mass density $\rho(>M_{\rm halo},z)$, given by~\cite{Boylan-Kolchin:2022kae}:
\begin{eqnarray}
\rho(>M_{\rm halo},z)=\int_{M_{\rm halo}}^\infty dM\,M\frac{dn(M,z)}{dM}\,.
\label{eq:rhohalo}
\end{eqnarray}
Finally, given values for the cosmic baryon fraction $f_b \equiv \Omega_b/\Omega_m$ and the efficiency of converting gas into stars $\epsilon \leq 1$, Eqs.~(\ref{eq:nhalo},\ref{eq:rhohalo}) straightforwardly translate into the cumulative comoving number density of galaxies more massive than $M_{\star}$ [$n_{\rm gal}(>M_{\star},z)$], the cumulative comoving mass density of stars contained in halos more massive than $M_{\rm halo}$ [$\rho_{\star}(>M_{\rm halo},z)$], and finally the cumulative comoving mass density of stars contained in galaxies more massive than $M_{\star}$: $\rho_{\star}(>M_{\star},z)$, which is the quantity we will be most interested in for what concerns the comparison to JWST observations. To be explicit, $\rho_{\star}(>M_{\star},z)$ is given by:
\begin{eqnarray}
\rho_{\star}(>M_{\star},z) = \epsilon f_b\rho(>M_{\rm halo},z) = \epsilon f_b\int_{M_{\rm halo}}^\infty dM\,M\frac{dn(M,z)}{dM}\,,
\label{eq:rhostar}
\end{eqnarray}
which directly follows from the identification $M_{\star}=\epsilon f_bM_{\rm halo}$. Clearly, a measured/inferred value $\rho_{\rm obs}$ of the cumulative comoving stellar mass density at a certain $z$ requires the underlying cosmology-dependent value of $\rho_{\star}(>M_{\star},z)$ to be at least as large as $\rho_{\rm obs}$ when $\epsilon=1$, i.e.\ in the most conservative and optimistic (but unrealistic) case~\cite{Menci:2022wia}.~\footnote{In principle we would need to correct the observed densities $\rho_{\rm obs}$ by a ``volume factor'' to account for the fact that the densities derived in Ref.~\cite{Labbe:2023ghw} explicitly assumed a $\Lambda$CDM cosmology, and similarly we would need to correct the measured masses by a ``luminosity distance factor'' to account for the fact that Ref.~\cite{Labbe:2023ghw} also inferred the latter from measured luminosities assuming $\Lambda$CDM. In practice, we have checked that both corrections are small for the values of the parameters considered here. Therefore, in the interest of carrying out an exploratory study, to zeroth order we neglect both corrections, which are subdominant to the effects of evolving DE and a nCC.} Finally, note that if one chooses to focus on a narrow redshift range $z \in [z_1;z_2]$ instead of a single (effective) redshift, Eq.~(\ref{eq:rhostar}) should be modified as follows:
\begin{eqnarray}
\rho_{\star}(>M_{\star},z) = \epsilon f_b\int_{M_{\rm halo}}^\infty dM\,M\frac{dn(M,z)}{dM}\frac{dV}{dz}\frac{dz}{V(z_1,z_2)}\,,
\label{eq:rhostarmodified}
\end{eqnarray}
where $V$ is the cosmic volume, such that $dV/dz=4\pi d_L^2/[H(z)(1+z)^2]$, with $d_L$ the luminosity distance, and $V(z_1,z_2)$ is the cosmic volume in the appropriate redshift range. For our purposes, we have checked that using Eq.~(\ref{eq:rhostarmodified}) with reasonable redshift limits in place of Eq.~(\ref{eq:rhostar}) makes little to no difference. Therefore, for simplicity we stick to Eq.~(\ref{eq:rhostar}) in what follows.

In what follows, we set our cosmological parameters (relevant to compute the expansion rate and matter power spectrum) to the following values: $H_0=67.32\,{\rm km}/{\rm s}/{\rm Mpc}$, $\Omega_m=0.3158$, $\Omega_b=0.049$, $n_s=0.96605$, and $\sigma_8=0.812$. For what concerns the parameters more of astrophysical interest, we set $f_b=0.156$, $\epsilon=0.32$, and $\delta_c=1.686$, unless otherwise stated. We have explicitly checked that reasonable excursions around these values have very little effect on our conclusions.

\section{Comparison to the JWST observations}
\label{sec:jwst}

Using $1-5\,\mu{\rm m}$ coverage data for galaxies identified by JWST NIRCam observations of the CEERS program, Ref.~\cite{Labbe:2023ghw} recently searched for intrinsically red, extremely massive galaxies at $7.5 \lesssim z \lesssim 10$, finding six candidate galaxies with stellar masses $\gtrsim 10^{10}M_{\odot}$. If the fiducial masses of these galaxies are confirmed, and most importantly their redshifts verified spectroscopically, the inferred cumulative comoving stellar mass density is of order $\rho_{\star}(>10^{10}M_{\odot}) \gtrsim 10^6M_{\odot}/{\rm Mpc}$ at $z \simeq 10$, significantly higher than anticipated from earlier studies based on rest-frame UV-selected samples~\cite{Stefanon:2022ghw}. Although issues related to calibration and the assumed initial mass function are still under debate~\cite{Lovell:2022bhx}, if confirmed, these early results are difficult if not impossible to realize within standard $\Lambda$CDM cosmologies. Our goal here is to examine whether an extended DE sector of the type discussed earlier, featuring a dynamical DE component in addition to a nCC, can potentially support the existence of this unexpectedly abundant population of extremely massive galaxies at very high redshift. We envisage this as being mostly an exploratory study, to work out whether such a possibility is even worth pursuing, and therefore shall limit ourselves to a zeroth order comparison between theory and observations, while leaving a significantly more detailed statistical analysis, whose goals would also include a stage of parameter inference, to follow-up work.
\begin{figure}[!t]
\includegraphics[scale=0.47]{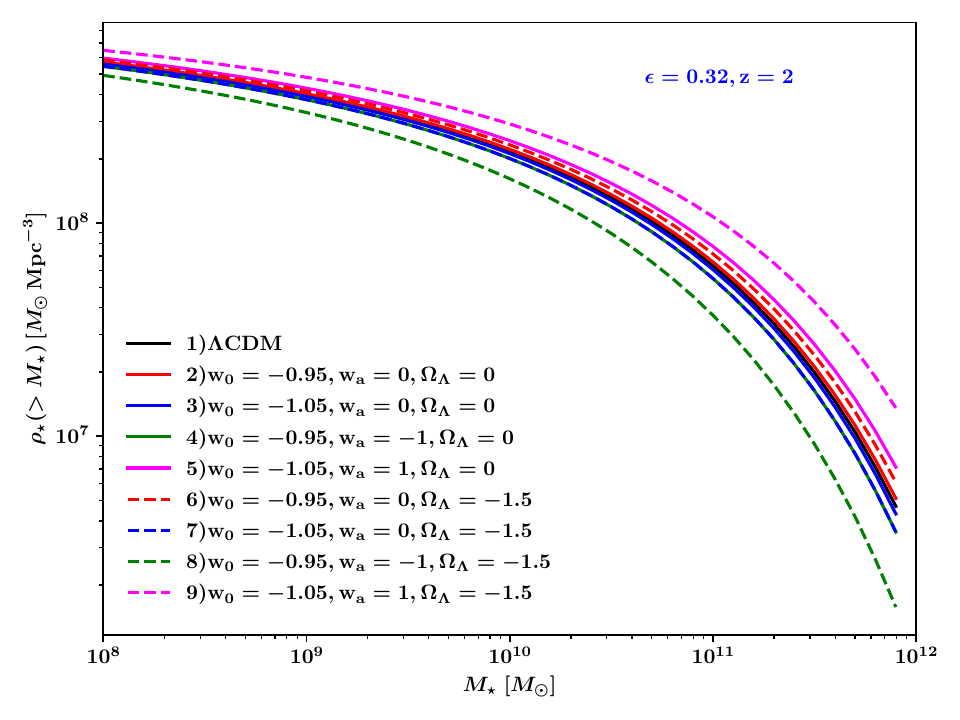}
\includegraphics[scale=0.47]{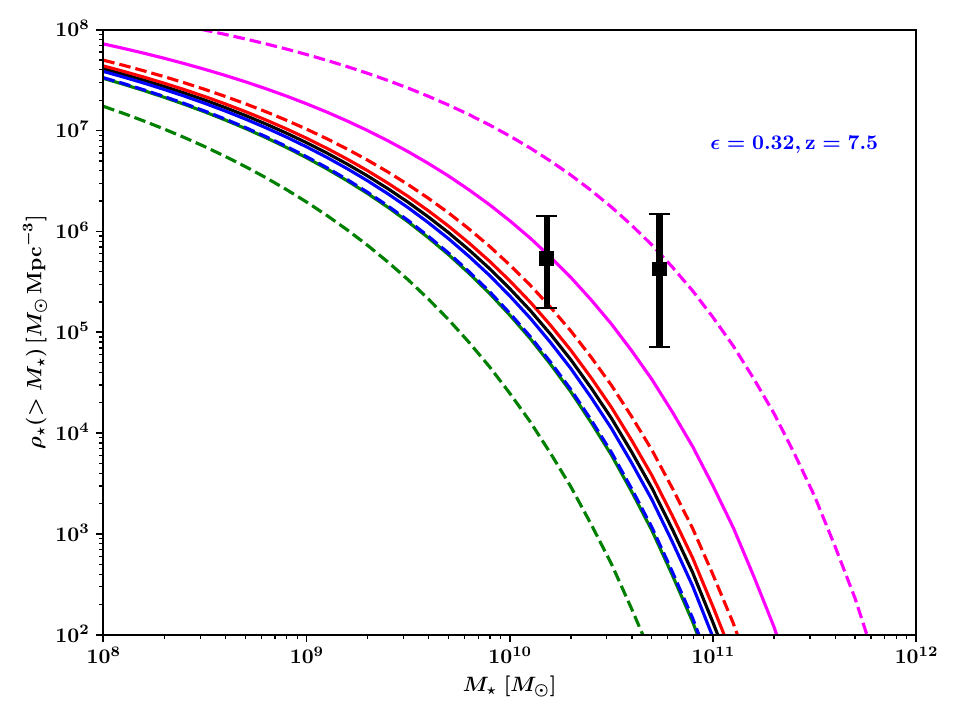}
\includegraphics[scale=0.47]{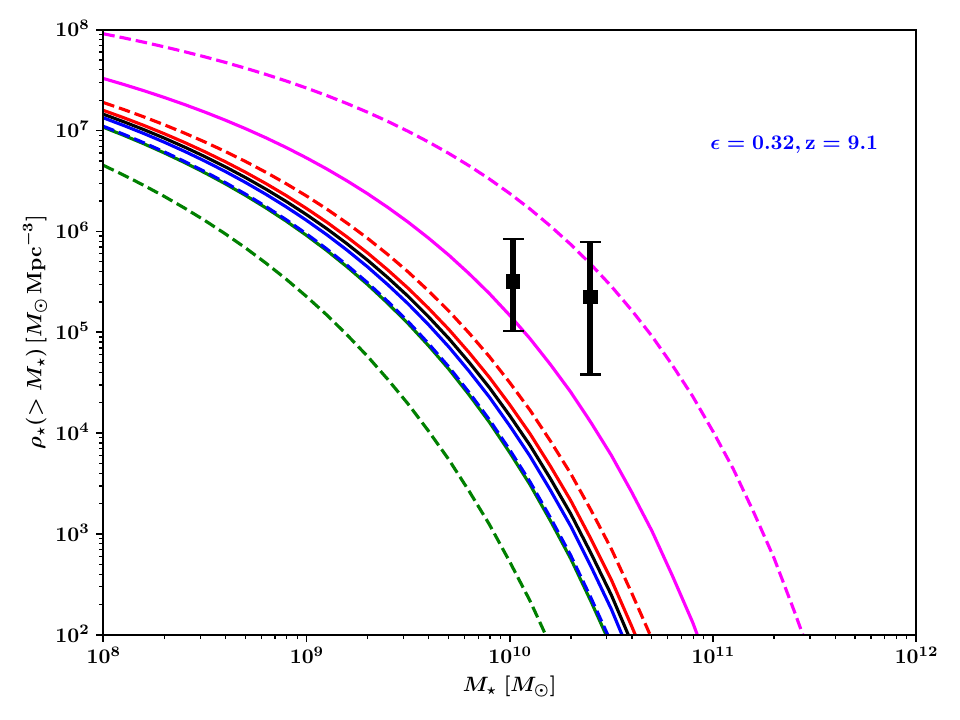}
\includegraphics[scale=0.47]{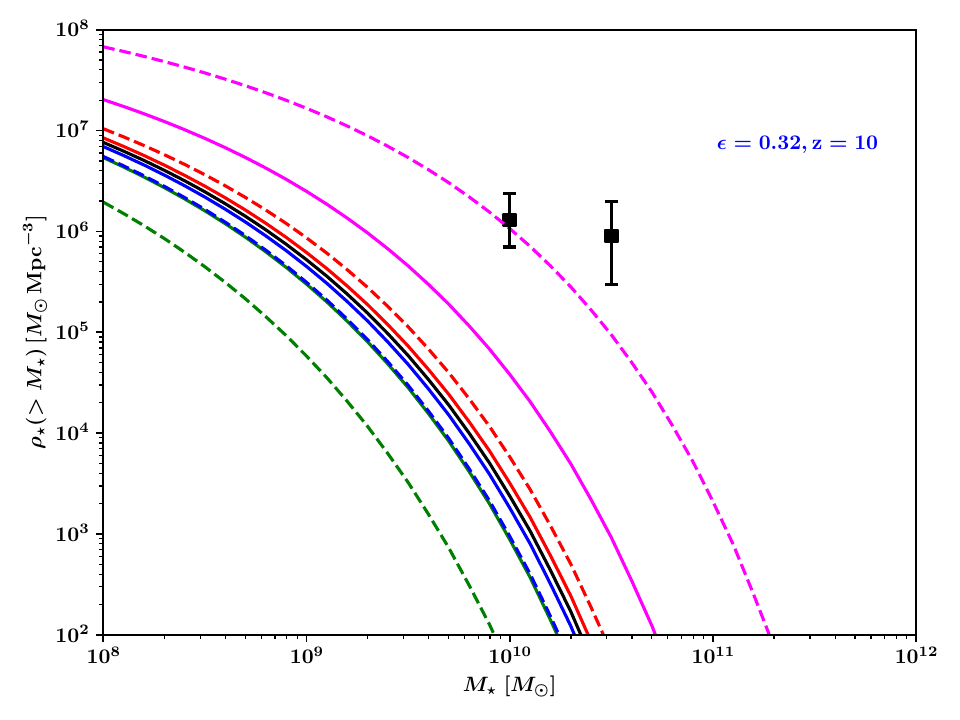}
\caption{Cumulative comoving stellar mass density as a function of stellar mass, $\rho_{\star}(>M_{\star},z)$, for the 9 different dark energy models summarized in Sec.~\ref{sec:models}, with the same color coding as in Fig.~\ref{fig:rhow}. The mass densities have been evaluated assuming a value of the efficiency of converting gas into stars $\epsilon=0.32$, and at redshifts $z=2$ (upper left panel), $z=7.5$ (upper right panel), $z=9.1$ (lower left panel), and $z=10$ (lower right panel). The datapoints within the three highest redshift bins are indicative of the cumulative stellar mass density inferred from the six most massive high-$z$ galaxies studied in Refs.~\cite{Labbe:2023ghw,Biagetti:2022ode}.}
\label{fig:rhostarm}
\end{figure}

We compute the cumulative comoving stellar mass density within galaxies with stellar mass above $M_{\star}$, $\rho_{\star}(>M_{\star})$, for the 9 different DE models discussed in Sec.~\ref{sec:models}, and for 4 representative redshifts: $z=2$, $z=7.5$, $z=9.1$, and $z=10$. In particular, the latter three redshifts are representative of the redshift range probed by the analysis of JWST NIRCam CEERS data in Ref.~\cite{Labbe:2023ghw}. In Fig.~\ref{fig:rhostarm} we show the theoretical predictions for $\rho_{\star}(>M_{\star})$ versus $M_{\star}$, alongside (for the three highest redshift bins) the cumulative stellar mass density inferred from the six most massive high-$z$ galaxies studied in Ref.~\cite{Labbe:2023ghw}, divided into three (effective) redshift bins.~\footnote{It is worth clarifying once more that we have not included the volume factor and luminosity distance factor corrections. Had we chosen to include them, it would not have been possible to produce Fig.~\ref{fig:rhostarm} in its current form, as each model would require a different correction factor, and therefore it would not have been possible to place all models on the same plot. However, we have checked that the effect of these corrections is small compared to the effects of evolving DE and the nCC.} From these plots, which we remark once more have been obtained assuming an efficiency of converting gas into stars $\epsilon=0.32$, high but not completely unrealistic, it is clear that these observations are in strong tension with the standard $\Lambda$CDM cosmology (given the assumed choice of cosmological parameters, in particular $\Omega_m$, $H_0$, and $\sigma_8$), whose predictions are represented by the solid black curve.

It is interesting to note that the differences between the different DE models considered, for what concerns the cumulative stellar mass density, are more noticeable at higher redshifts. In particular, as we will discuss in more detail shortly, for some of the DE models considered and in certain mass ranges, $\rho_{\star}$ can be up to a couple of orders of magnitude larger compared to the $\Lambda$CDM prediction. This is in agreement with the earlier results shown in Fig.~\ref{fig:growthfactor} and Fig.~\ref{fig:percent}, in turn reflecting the fact that the growth factor $D(z)$, which is the quantity relevant for computing the linear power spectrum $P(k)$ and hence the halo mass function $dn/dM$, is normalized to $\delta(z=0)$. Our results therefore confirm that observations of massive, high-$z$ galaxies (such as those detected by JWST) can potentially shed light on the dynamics of DE, an intrinsically late-time phenomenon, through its indirect imprint on the growth of structure at much higher redshift.

In agreement with the earlier results shown in Fig.~\ref{fig:growthfactor} and Fig.~\ref{fig:percent} and discussed in Sec.~\ref{subsec:densityperturbationsgrowth}, we find that the model leading to the most significant departures from $\Lambda$CDM is model 9), represented by the dashed magenta curve, where besides a nCC the DE sector consists of a dynamical DE component which is phantom today but was quintessence-like in the past, and hence crossed the phantom divide at some point during its evolution. For the particular choice of parameters made in this work, which we stress once more is made purely for exploratory purposes, the phantom divide is crossed when the scale factor of the Universe is $a_p=(w_0+w_a+1)/w_a = 0.95$, or equivalently at redshift $z_p = 1/a_p-1 = -(w_0+1)/(w_0+w_a+1) \approx 0.05$.

As we see from the panels corresponding to the highest redshift bins in Fig.~\ref{fig:rhostarm}, at the masses relevant for the JWST observations the theoretical predictions for the cumulative stellar mass density in these dynamical DE+nCC cosmologies can be up to four orders of magnitude larger compared to $\Lambda$CDM, and are in good agreement with the JWST observations, with the only datapoint which is more than $1\sigma$ off being that corresponding to the highest mass bin at $z=10$ (rightmost datapoint in the lower right panel). Moreover, as is clear from the same panels (compare the dashed versus solid magenta curves), the inclusion of the nCC is crucial in order to increase the theoretical predictions for the cumulative stellar mass density, thereby improving agreement with the JWST results. As discussed earlier, the reason is that including a nCC (and hence a larger, more negative $\Omega_{\Lambda}$) requires increasing the positive energy density of the dynamical DE component $\Omega_x$ in order for $\Omega_{\rm DE} \approx 0.7$ to hold, in agreement with cosmological observations.

The final interesting point to note is that the combination of phantom DE and a nCC has been shown to possess interesting features in light of the Hubble tension~\cite{Dutta:2018vmq,Visinelli:2019qqu,Sen:2021wld}. In fact, due to the well-known negative correlation between $w$ and $H_0$, phantom DE and/or effective phantom components (e.g.\ arising from modified gravity) can help raise $H_0$ and therefore partially alleviate the Hubble tension, as investigated in a large number of works, see e.g.\ Refs.~\cite{DiValentino:2016hlg,Zhao:2017cud,Mortsell:2018mfj,Li:2019yem,Vagnozzi:2019ezj,DiValentino:2019ffd,Dutta:2019pio,DiValentino:2019jae,Zumalacarregui:2020cjh,Alestas:2020mvb,Yang:2020ope,Alestas:2020zol,Kumar:2021eev,Teng:2021cvy,Krishnan:2021dyb,Bag:2021cqm,Theodoropoulos:2021hkk,Alestas:2021luu,Roy:2022fif,Heisenberg:2022lob,Chudaykin:2022rnl,Akarsu:2022typ,Santos:2022atq,Schiavone:2022wvq,Ballardini:2023mzm,Zhai:2023yny,Montani:2023xpd,Escamilla:2023oce,Cline:2023cwm,Vagnozzi:2023nrq}.~\footnote{Conversely, it is known that the simplest quintessence models based on a single scalar field with canonical kinetic term, minimally coupled to gravity, and in the absence of higher derivative operators, are marginally disfavored observationally, as they worsen the Hubble tension~\cite{Vagnozzi:2018jhn,OColgain:2018czj,Colgain:2019joh,Banerjee:2020xcn,Heisenberg:2022gqk}.} These features can be further enhanced in the presence of a nCC, as shown for instance in Refs.~\cite{DiValentino:2020naf,Adil:2023exv} in the context of DE models for which phantom crossing and negative DE energy densities are allowed, particularly for models where the phantom divide is crossed going from quintessence-like DE at early times to phantom DE today. This makes the combination of phantom DE and a nCC an interesting one in light of the Hubble tension, at least from the phenomenological point of view. An important clarification is however in order. It is known that phantom DE, and more generally late-time new physics, can never fully solve the Hubble tension \textit{alone} while remaining in agreement with Baryon Acoustic Oscillations and Hubble flow Type Ia Supernovae data~\cite{Bernal:2016gxb,Addison:2017fdm,Lemos:2018smw,Aylor:2018drw,Schoneberg:2019wmt,Knox:2019rjx,Arendse:2019hev,Efstathiou:2021ocp,Cai:2021weh,Keeley:2022ojz}. Therefore, even though the phantom+nCC model is interesting in this regard, we know that it cannot be the end of the story, at least if we take these late-time datasets seriously. Nevertheless, we remark that a DE model displaying phantom nature at late times and quintessence-like nature at early times (with phantom crossing in between), supplemented with a nCC in the dark sector [e.g.\ our model 9) with $w_0=-1.05$, $w_a=1$, $\Omega_\Lambda=-1.5$, and $\Omega_x=2.2$], remains a phenomenologically very interesting one, given its ability to simultaneously alleviate the Hubble tension (in part) and explain the puzzling JWST observations of high-$z$ galaxies. We thus believe that this class of models is worthy of further studies.

\section{Conclusions}
\label{sec:conclusions}

The first glimpse of high-$z$ galaxy formation from the James Webb Space Telescope (JWST) has delivered extremely surprising results which, if confirmed spectroscopically and by other means, have the potential to revolutionize our understanding of the Universe. In particular the surprising abundance of massive, high-$z$ galaxies observed by JWST appears to pose a serious challenge for the $\Lambda$CDM model and, in turn, can be used to test alternative cosmological models which allow for a faster growth of structure at early times. Our work falls within the promising and fruitful direction of testing new fundamental physics using high-$z$ galaxies:~\footnote{See also Refs.~\cite{Jimenez:2019onw,Vagnozzi:2021tjv,Valcin:2021jcg,Bernal:2021yli,Cimatti:2023gil} for related works in this direction.} in particular, we have further strenghtened the case for the use of the abundance of high-$z$ galaxies inferred from JWST data to test dark energy (DE) models, complementing standard tests of DE which are typically based on probes at very low redshift.

We have considered a DE sector featuring a negative cosmological constant (nCC) and a time-varying DE component with positive energy density, modelled following the CPL parametrization, and allowed to cross the phantom divide $w_x=-1$. Such a scenario carries strong motivation from string theory, where AdS vacua appear ubiquitously, while the construction of stable dS vacua has so far proven extremely hard~\cite{Danielsson:2018ztv}. We have shown that the combination of phantom-crossing DE (with equation of state $w_x>-1$ in the past and $w_x<-1$ today) and a nCC can enhance structure growth at early times, potentially bringing the predicted cumulative comoving stellar mass density in agreement with the JWST observations, an agreement which cannot be achieved within the standard cosmological model even if star formation is maximally efficient ($\epsilon=1$). It is worth noting that such a model also has the ability to alleviate (while not fully solving) the Hubble tension~\cite{Visinelli:2019qqu,DiValentino:2020naf,Sen:2021wld}. We believe this combination of ingredients is therefore interesting from all three the observational, phenomenological, and theoretical (string-inspired) perspective.

While our work should be intended as an exploratory analysis, it reinforces the enormous potential held by observations of the abundance of high-$z$ galaxies in probing new fundamental physics, including string theory-inspired ingredients. These considerations alone are reason enough to keep exploring this promising direction from multiple related perspectives, one of which could be a first-principles implementation of well-motivated string scenarios, going beyond the phenomenological parametrization we have adopted, as done for example in Refs.~\cite{Cicoli:2018kdo,Ruchika:2020avj}. Another interesting direction to pursue would be that of performing a complete statistical/parameter inference analysis in light of JWST observations (which would require a more careful treatment of the volume and luminosity distance factor corrections), as well as a more detailed study of the complementarity between the latter and standard cosmological probes~\cite{SimonsObservatory:2018koc,SimonsObservatory:2019qwx}. In particular, such an analysis would allow us to constrain the DE parameters $w_0$, $w_a$, and $\Omega_{\Lambda}$ from a joint analysis of JWST data and standard cosmological data, allowing us to assess the degree to which the nCC scenario we have studied is preferred (if any) over $\Lambda$CDM. All of these directions are potentially extremely fruitful, and we plan to report on these and related results in future work.

\section*{Acknowledgments}
We are grateful to Chethan Krishnan, Nicola Menci, Eoin \'{O} Colg\'{a}in, Ruchika, and Shahin Sheikh-Jabbari for useful discussions. A.A.S. and U.M. acknowledge support from the Science and Engineering Research Board (SERB) of the Government of India through research grant no.~CRG/2020/004347. S.A.A., U.M., and A.A.S. acknowledge the use of the High Performance Computing facility Pegasus at IUCAA, Pune, India. S.V. acknowledges support from the Istituto Nazionale di Fisica Nucleare (INFN) through the Iniziativa Specifica ``FieLds And Gravity'' (FLAG). This publication is based upon work from COST Action CA21136 – ``Addressing observational tensions in cosmology with systematics and fundamental physics (CosmoVerse)'', supported by COST (European Cooperation in Science and Technology).

\bibliographystyle{JHEP}
\bibliography{JWSTnegativeCC}

\end{document}